\documentclass[prb,twocolumn,showpacs,preprintnumbers,amsmath,amssymb]{revtex4}

\usepackage{graphicx}
\usepackage{dcolumn}
\usepackage{bm}

\begin{document}

\title{Electrical transport properties of stable single-atom contacts of zinc}

\draft

\author{A. Mayer-Gindner,$^1$ H. v.
L\"ohneysen,$^{1,2}$ and E. Scheer$^3$}

\affiliation{$^1$ Physikalisches Institut, Universit\"at
Karlsruhe, D-76128 Karlsruhe, Germany\\
$^2$ Forschungszentrum Karlsruhe, Institut f\"ur
Festk\"orperphysik,
           D-76021 Karlsruhe, Germany\\
$^1$ Fachbereich Physik, Universit\"at Konstanz,
           D-78457 Konstanz, Germany}
\date{\today}

\begin{abstract}
  We report low-temperature measurements of the electrical transport properties
 of few-atom contacts of the superconducting metal zinc,
 arranged with lithographically fabricated mechanically controllable breakjunctions (MCB).
The conductance histogram shows several narrow peaks, not
regularly observed for multivalent metals. The first peak -
corresponding to the single-atom contact - is located slightly
below one conductance quantum and is split into two subpeaks,
possibly indicating two preferred configurations of the
single-atom contact. The current-voltage characteristics in the
superconducting state show nonlinearities due to multiple Andreev
reflections (MAR). In addition we find nonlinearities in the
current-voltage characteristics
 that we attribute to electronic-interaction effects due to the granular structure of the film.
 Some of these latter findings are in qualitative agreement with a prediction by Avishai et
 al. \cite{avishai}.

\end{abstract}
\pacs{PACS numbers: 73.40.Jn, 74.50.+r, 73.20.Dx}

\maketitle

 For
revealing and understanding the electronic transport properties of
atomic scale circuits, single- or few-atom contacts are used as
model systems. In this limit the electronic conduction can be
regarded as a fully quantum-mechanical scattering and
wave-matching problem, i.e. the conductance can be attributed to
independent electronic modes, nick-named "conduction channels".
 An atomic-size contact
between two metallic electrodes can accommodate only a small
number of conduction channels. The transport properties are thus
fully described by a set $\left\{ \tau_{\mathrm n}\right\}
=\left\{\tau_1,\tau_2,...\tau_{\mathrm N}\right\} $ of
transmission coefficients which depends both on the chemical
properties of the atoms forming the contact and on their
geometrical arrangement. Experimentally, contacts consisting of
even a single atom have been obtained using both scanning-tunnel
microscopy and breakjunction techniques \cite{agrait03}. The total
transmission $D$=\thinspace $\sum\limits_{i=1}^N \tau_{\mathrm i}$
of a contact is deduced from its conductance $G$ measured in the
normal state, using the Landauer formula $G~=~G_0D$ where
$G_0=2e^2/h$ is the conductance quantum \cite{landauer70}.

Experiments on a large ensemble of metallic contacts have
demonstrated the {\it statistical} tendency of atomic-size
contacts to adopt element-specific preferred values of
conductance. The actual preferred values depend not only on the
metal under investigation but also on the experimental conditions.
However, for many metals,
 and in particular 'simple' ones (like Na, Au...)
which in bulk form good 'free electron' metals, the smallest
contacts have a conductance $G$ close to $G_0$ (Ref.
\cite{agrait03}). Statistical examinations of Al point contacts at
low temperatures yield preferred values of conductance at $G =
0.8~G_0, 1.9~G_0, 3.2~G_0$ and $4.5~G_0$ (Ref.
\cite{yansoncomment}), indicating that single-atom contacts of Al
have a typical conductance slightly below the conductance quantum.

However, it has been shown by the analysis of current-voltage (IV)
characteristics in the superconducting state that Al single-atom
contacts accommodate in general three transport channels, the
transmissions of which add up to a value around 1 (Ref.
\cite{aluprl}). These findings are in agreement with measurements
of the shot noise \cite{brom,cron}, conductance fluctuations
\cite{ludoph1}, thermopower \cite{ludoph2} and the supercurrent
\cite{goffman} in such contacts. A quantum chemical model
\cite{nature} which links the $\tau_{\mathrm n}$ to the chemical
valence and the atomic arrangement of the region around the
central atom of the constriction, has been put forward. This model
predicts that monovalent metals would transmit a single channel
while sp-like metals  as e.g. Al or Pb up to four channels due to
the contributions of the (unoccupied) p orbitals.
 So far Al appears to
be the only metal with the property to show peaks in the histogram
close to multiples of $G_0$ although this does not correspond to
quantized transmission values.

According to simplistic arguments and ignoring the band structure,
divalent elements such as the alkaline-earth elements or the IIB
subgroup elements Cd, Zn and Hg, should be insulating, since they
have a completely filled outer s-shell. This apparent discrepancy
to the experimental findings indicates that additional orbitals
besides the s-orbitals contribute to the electronic conductance.
Tight-binding calculations reveal that in the case of Zn the 4p
orbitals are the most relevant ones for a correct description of
the bulk band structure (Ref. \cite{papaconstantopoulos}). With
these considerations, single-atom contacts of Zn are expected to
have very similar transport properties as Al single-atom contacts,
i.e. up to four channels with a total conductance of about 1~$G_0$
\cite{nature}. The distribution of the transmission coefficients,
however, may deviate, since it depends on the exact atomic
configuration. Since Zn crystallizes in a hexagonal, slightly
distorted hcp  structure while Al is a fcc crystal, the actual
atomic arrangement of the point contacts might be different.
Moreover, due to the "loss" of many neighbors, a single Zn atom in
a contact may behave more atomic-like.

Here, we report the fabrication and low-temperature electronic
transport properties of lithographic MCBs of Zn. This work
represents the first study of atomic-size contacts of Zn. The
measurements in the normal state demonstrates the tendency of the
atomic-size contacts to adopt preferred conductance values. The
analysis of the current-voltage characteristics reveals
superconducting properties of the nanowires as well as indications
of electron-electron interactions. We show data of four, nominally
identical samples, labeled No.~1 to 4.

\begin{figure}[t]
 \begin{center}
\includegraphics[width=0.8\columnwidth,clip]{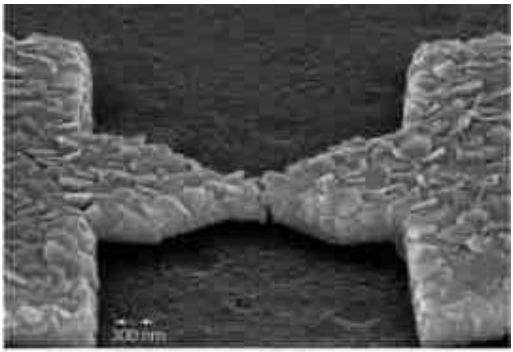}
    \caption{\label{amg_f1} Electron micrograph of sample No. 1
    under an inclination angle of 70$^{\rm o}$, taken after measurement.
    The poly-crystalline structure of the Zn film is clearly visible.
    A remaining spacing of the bridge arms of several nm and a fine
     nanowire between them is observed.}

    \end{center}
\end{figure}

The samples are fabricated along the lines of Ref.
\cite{jansaclay}. Our samples are 2\thinspace $\mu $m long, 200 nm
thick suspended
Zn nanobridges, with a $200$\thinspace ${\rm nm}\times 100$\thinspace ${\rm %
nm}$ constriction in the center (cf.~Fig.~\ref{amg_f1}). The
thickness of the bronze substrates is $\approx 0.3$~mm, the
distance between the counter-supports of the bending mechanism is
10~mm. With these sample parameters and the geometry of the
mechanics we achieve reduction ratios (between the motion of the
pushing rod and the displacement of the bridge arms) of the order
of 5000:1 (see below).

Special care has to be given to the evaporation of Zn. Because of
its low boiling temperature and high vapor pressure, Zn desorbs
easily from the substrate if no crystallization seeds are
provided. If Zn is evaporated at room temperature, it grows in
islands of micron-size with the hexagonal axis preferably
perpendicular to the substrate plane. Since this structure is not
suitable for the formation of atomic-size contacts, we take
special care to reduce the grain size. We first deposit a 1~nm
thick seed layer of Ge onto the substrate which is cooled by
liquid nitrogen to a temperature of $\approx -50^{\rm o}{\rm C}$.
At lower substrate temperatures the different thermal expansions
of the metallic substrate, the Zn itself, the polyimide
sacrificial layer and the polymer lithographic mask provoke cracks
in either the mask or the Zn layer on warming up to room
temperature. The metal is then evaporated from a home-made
boron-nitride crucible heated by a tungsten foil at a pressure of
$\approx 10^{-6}$ mbar and a rate of $\approx 0.2$~nm/s at the
same substrate temperature. The parameters are chosen empirically
such that the grain size of the Zn layer does not exceed $\approx
200$~nm and the adhesion is good enough for the subsequent
lift-off process. Finally, the bridge is underetched in an
isotropic oxygen plasma which reduces the height of the polyimide
layer by about 300 to 500~nm. At the narrow constriction the
nanobridge is now unsupported. The resistance of the nanobridge
does not increase during the etching process, indicating that
reaction between the oxygen ions and Zn is weak.

The bridge is then mounted on a three point bending mechanism on a
stick which is immersed into liquid helium and can be pumped down
to reach temperatures of 1.5~K.
 The residual resistivity ratio of the bridges is
typically RRR = $R(300{\rm K)}/R(1.5{\rm K})\thinspace \simeq
\thinspace 2.4$ from which we deduce an elastic mean free path of
13 nm. On an unpatterned thin film of the same thickness we
measured the critical temperature $T_{\mathrm c} \thinspace =
\thinspace 0.78$\thinspace K and a perpendicular critical field of
$B_{\mathrm c} \thinspace = \thinspace 12$\thinspace mT.
$T_{\mathrm c}$ is slightly reduced with respect to the bulk
critical temperature of $T_{{\mathrm c},bulk} = 0.89$~K while
$B_{\mathrm c}$ is enhanced, as usual in thin films ($B_{{\mathrm
c},bulk} = 5 $~mT) (Ref. \cite{kittel}).

As a first step we have determined the preferred conductance values of few-atom contacts
when opening or closing continuously and have constructed the corresponding conductance
histograms.

A screw with 500\thinspace $\mu $m
pitch, driven by a dc-motor on top of the cryostat
through an (exchangeable) reduction gear box,
controls the motion of the pushing rod that bends the substrate.
The relative displacement of the pushing rod can be controlled to
a precision of $\approx \thinspace 4\thinspace \mu$m, which due to
the geometry of the bending mechanism results in a relative motion
of the two anchor points of the bridge of around 0.9\thinspace pm.
This was verified using the exponential dependence of the
conductance on the inter-electrode distance in the tunnel regime.
The absolute precision is restricted due to hysteresis of the
bending mechanism (torsion of the pushing rod etc.) and roughness
of the screw pitch to about 100~nm for the pushing rod
(corresponding to 0.02~nm for the bridge). We note that the
stability of the setup allows us to halt at any point along the
breaking or closing curves, even on the last plateau when the
contact is formed by a single atom, without any observable drift
of the measured conductance.

The resistance of the sample is recorded by a 4-point resistance
bridge with a measuring current of 1~nA and a resolution of
$10^{-3}$ in a range from 200~$\Omega$ to 200~k$\Omega$. Due to
the mentioned texture of the evaporated film there are only few
sliding planes parallel to the substrate which would allow for a
stretching of the film. Consequently it often happens that the two
electrodes remain separated  by a gap of several nm after the
first breaking. Therefore the first opening of the contact has to
be performed very slowly. Typical opening speeds for the
systematic measurements are 155 nm/s for the pushing rod,
corresponding to a displacement of the bridge anchor points of 36
pm/s. The output of the resistance bridge is recorded by an
oscilloscope and then transferred to a computer for calculating
the histogram.

\begin{figure}[t]
 \begin{center}
\includegraphics[width=0.95\columnwidth,clip]{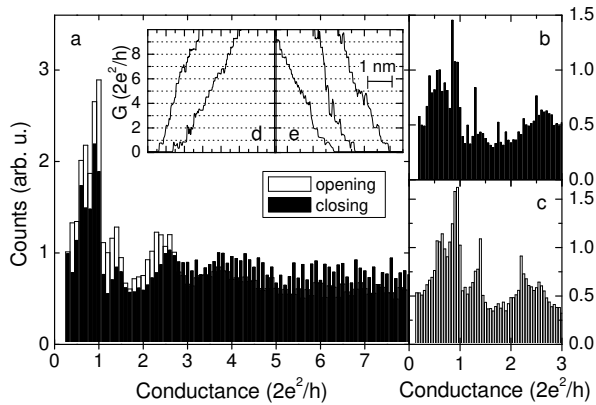}
    \caption{\label{amg_f2} a (b and c): Conductance histograms for opening and closing the contacts,
    recorded on sample No. 1 (4) at $ T = 1.5$~K. The histograms have been calculated
     from $\approx 600$ individual opening and closing curves, respectively. d (e)
     Examples for typical closing (opening) curves.}
\end{center}
\end{figure}

The conductance $G$ decreases in steps of the order of 0.2 to
1\thinspace $G_0$, their exact sequence changing from opening to
opening (see panel e of Fig.~\ref{amg_f2}). Since the mechanical
deformation is different for opening and closing, we construct
separate histograms for both experiments. Fig. \ref{amg_f2}a
displays the conductance histograms calculated from $\approx 600$
subsequent opening and closing sequences recorded during two weeks
on sample No. 1 which has been kept at low temperatures. Similar
histograms were achieved with a lower motion speed of the bridge
arms of 17 pm/s. After an accidental warming up to $\approx 80$~K
no step curves could be observed any more.

Several peaks appear for both motion directions at similar
positions of $0.8\thinspace G_0$, 1.4\thinspace $G_0$ and $\approx
2.5~G_0$. Interestingly the first peak is split into two
substructures at $\approx 0.7~G_0$ and 0.9~$G_0$, a feature which
has not been observed for other materials. This splitting has been
observed in several independent measurements carried out with two
different samples (Nos. ~1 and 4). Although the splitting is
observed in all our measurements, the absolute values of the peak
positions and the relative peak heights vary slightly from
experiment to experiment (see Fig.~\ref{amg_f2}b and c).  A
possible explanation of the splitting would be two different
configurations of the central atom e.g. different crystalline
orientations or a short contact with only one plane with a single
atom in cross-section vs. a dimer contact with two single atoms in
series. The latter configuration has been found in molecular
dynamics calculations to be typical for Au contacts just before
breaking \cite{dreher}. This geometrical difference gives rise to
different conductances \cite{haefner}. In similar experiments by
Yanson et al. \cite{yansondiss} on "classical" breakjunctions made
of notched Zn wires, no splitting of the first peak is found, but
a shoulder that is compatible with our findings. The peaks at
higher conductance values are in reasonable agreement with our
findings as well. It is well known that details of the histograms
do depend on the experimental conditions, as one can easily see by
comparing different published results of e.g. the most extensively
studied metal gold. In the experiment with Zn by Yanson et al.
higher voltages have been applied, a slightly higher temperature
and faster opening speeds have been used and only opening traces
were recorded.

In our experiment, the main difference between opening and closing
consists in the peak heights. The low-conductance peaks are less
well pronounced when closing. In the closing traces (see panel d)
we often observe that the first contact has a conductance well
above $G_0$. We attribute this to the formation of contacts larger
than a single-atom contact since the apex atoms from the previous
opening sequence might have relaxed back to positions deeper in
the tip.

Conductance histograms of multivalent metals in general do not
show well pronounced multiple peaks
\cite{yansondiss,ludoph3,costa-kramer}. An exception of this
general observation was so far Al, as mentioned already in the
introduction. The histogram of Zn shows similarities to that of
Al: Several peaks occur with an average spacing between the peaks
of about 1.3~$G_0$. One could argue whether this similarity in the
preferred conductance values would be mirrored in the quantum
transport properties, i.e. the conduction channels.
Therefore we
performed measurements in the superconducting state for the
determination of the channel ensemble. For this purpose extremely
stable contacts and thus a more complex bending mechanism is
required. The samples are mounted onto a three-point bending
mechanism thermally anchored to the mixing chamber of a dilution
refrigerator and that contains now a differential screw close to
the sample holder. The pushing rod of the bending mechanism is
driven by a dc motor at room temperature via a mechanical
feedthrough and several gear boxes with a reduction ratio of
1:2,400,000 between the turning speed of the motor and the axis
entering the differential screw. The differential screw transforms
one turn of the axis into a motion of the pushing rod of
100~$\mu$m. The sample geometry is the same as in the previous
measurements, giving again rice to a reduction ratio of 1:5000.
Taking into account the reduction of the gear boxes, the
differential screw and the geometry of the sample itself, one turn
of the motor yields a displacement of the bridge arms of the
sample of about 8~fm. Therefore, stable tunnel or few-atom wide
contacts are achievable, but no histograms can be recorded within
a reasonable time. For further experimental details see Refs.
\cite{aluprl} and \cite{physicab}. Contrary to the experiments
described in \cite{aluprl}, only lossy coaxial wires and
twisted-pair wires in capillaries were used for high-frequency
filtering. This restricts the electronic temperature of the
samples to about 160~mK.

\begin{figure}[t]
  \begin{center}
\includegraphics[width=0.8\columnwidth,clip]{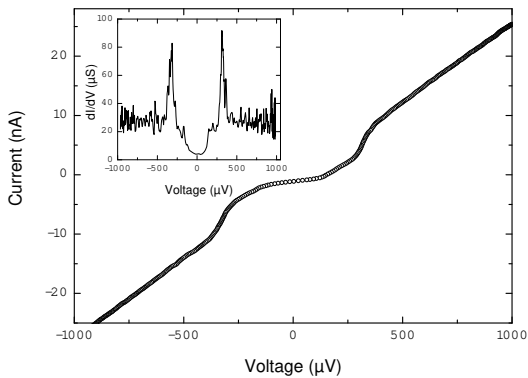}
    \caption{\label{amg_f3} Current-voltage characteristic (IV) of sample No. 2 when in the
near tunneling regime at $T = 140$~mK.
    From the slope at voltages higher than the superconducting gap we deduce a conductance of $0.34~G_0$.
    The superconducting gap is $\Delta \simeq 165~\mu $eV. Inset: Numeric derivative of the IV. }

    \end{center}
\end{figure}

Fig. \ref{amg_f3} depicts a current-voltage sweep (IV) measured at
$T = 160~$mK when sample No.~2 was in the close tunnel regime with
a conductance of $G = 0.34~G_0$. In the inset, the differential
conductance is shown. The tunnel contact was achieved by carefully
closing the contact right before the so-called "jump to contact",
i.e. the strong conductance increase signalling hte formation of
an atomic bond between the two bridge arms \cite{agrait03}.
Apparently, the spectrum deviates from the expected behavior for
BCS superconductors, as has already been observed in earlier
studies of tunnel contacts of Zn fabricated by different methods
\cite{tomlinson}. We deduce a superconducting gap $\Delta =
165~\mu e $V, somewhat higher than the values found in the
literature, which vary typically between $\Delta = 120~\mu$eV and
$ 145~\mu$eV, depending on the measuring method, crystallographic
orientation and other parameters \cite{tomlinson}. After these
characterization measurements, we close the bridge again to form
single- or few-atom contacts. This is done by reducing the bending
of the substrate until the sample achieves a conductance of
$\approx 10~G_0$ and opening it again carefully. Just before break
typically conductances of $0.7-1.4~G_0$ are observed.

\begin{figure}[t]
  \begin{center}
\includegraphics[width=0.8\columnwidth,clip]{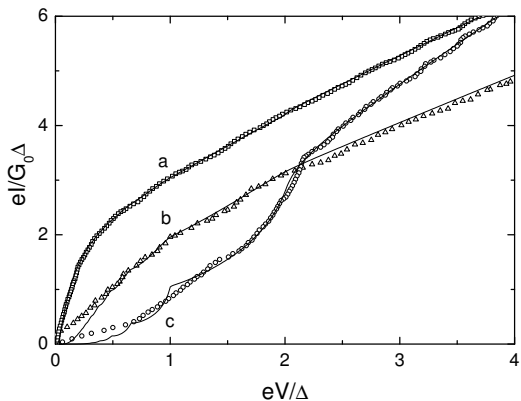}
    \caption{\label{amg_f4} Current-voltage characteristics of few-atom contacts of
    sample No. 2.
    The measuring temperature was $T \simeq 160$~mK. The symbols
    are experimental data, while the fits are calculated
    with the theory of MAR from Refs.~[\onlinecite{averin,carlosprb}] for
    $\Delta = 165~\mu e$V and $T = 160$~mK.
    The transmission coefficients are: Curve (a)
$\tau_1 = 0.97,~\tau_2 = 0.03$; (b) $\tau_1 = 0.84$; (c) $\tau_1 =
0.69,~\tau_ 2 = 0.39,~\tau_3 = 0.23,~\tau_4 = 0.15.$ }

    \end{center}
\end{figure}

 In Fig. \ref{amg_f4}
we depict a typical set of IVs recorded on contacts on last
plateaus of sample No. 2 before breaking of the wire, measured at
temperatures around 160 mK. The typical nonlinearities due to
multiple Andreev reflection (MAR) are observable
\cite{averin,carlosprb}. However, they appear rounded compared to
the theoretical prediction and the observations for contacts of
superconductors with a density of states obtained from ${\rm
d}I/{\rm d}V$ vs. $V$ that is closer to the BCS shape. This fact
limits the precision with which the channel ensemble can be
determined. The lines shown in Fig.~\ref{amg_f4} are fits to the
theory of MAR taking into account 1, 2 or 4 channels for curves a,
b, and c (all measured on sample No. 2). Because of the rounded
MAR structures, only the number of channels can be estimated, but
no meaningful value for the transmissions can be given for
contacts with more than one channel because several channel
transmissions combinations give fits with the same quality. The
fact that the determination of the number of channels is more
robust than the precision with which the individual transmissions
can be deduced has already been reported for few-atom contacts of
Nb, a metal that also shows a DOS deviating from the BCS shape
\cite{ludoph3}. The robustness of the determination of the number
of channels is due to the super-linear dependence of the excess
current (the extrapolated intersection of the high-bias current
with the current axis) on the transmission values
\cite{averin,carlosprb}. Interestingly, contacts with a dominating
single channel, accompanied by a second, but much smaller one,
similar to the observations for Au \cite{agrait03,scheer01}, are
achievable. These IVs are routinely observed at the end of a last
plateau after an abrupt change of the IV including an increase of
the conductance, and thus an re-arrangement of the atomic-size
contact. Once such a contact has been established, it remains
stable for an elongation of the bridge of 1-2~\AA . For Al
samples, contacts with a dominating channel with the contribution
of one or two smaller channels could be observed as well
\cite{aluprl,goffman}. However, for Al these contacts evolved
continuously from a situation with three channels of similar size
and occurred only very rarely \cite{boehler}.

The origin of these single-channel like contacts and their
possible correlation with the observed splitting of the first
maximum in the conductance histogram has to be clarified further.
Therefore, measurements on contacts, fabricated from whisker
breakjunctions have been carried out. The results of these
investigations together with a detailed theoretical analysis will
be reported elsewhere \cite{haefner}. We expect the
superconducting DOS to be closer to the BCS DOS for those samples
and consequently allowing to determine the transmission values
precisely.

Another observation hampers the quantitative determination of $\{
\tau_n \}$. As can be seen in curve c of Fig. \ref{amg_f4}, the
current onsets corresponding to MAR processes involving two or
more electrons seem to be shifted to higher voltages with respect
to the current onsets predicted by the theory and observed for
contacts of other superconductors \cite{nature}. We attribute this
observation to the onset of Coulomb blockade \cite{grabert} due to
the granularity of the film. Our findings are in qualitative
agreement with a prediction by Avishai et al. \cite{avishai}. The
authors have calculated the IV characteristics of an interacting
quantum dot between two superconductors and found a decrease of
the subgap current and a shift of the higher-order MAR processes
to higher voltages. However, since no quantitative theory for our
experimental realization and resulting transmission coefficients
are available, we cannot perform a quantitative comparison with
this suggestion.

Cron et al. \cite{crondiss} have placed an Al breakjunction in a
high-impedance environment in order to study the dynamic Coulomb
blockade \cite{grabert} of a single junction. They found a
reduction of the low-bias conductance in reasonable agreement with
a theoretical model when the contact was driven normalconducting
by applying a small magnetic field. However, no effect was
detected in the superconducting state.

\begin{figure}[t]
   \begin{center}
\includegraphics[width=0.8\columnwidth,clip]{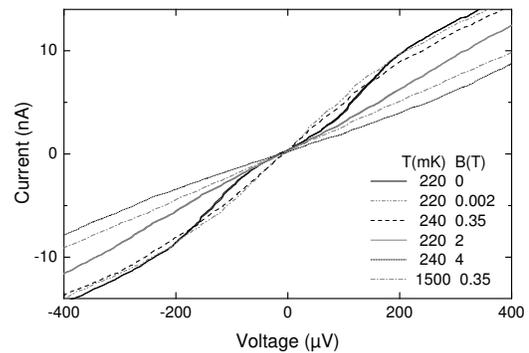}
    \caption{\label{amg_f5} Current-voltage characteristics of a tunnel
    contact of sample No. 3 with a conductance  of $\sim 0.3~G_0$ for the
    magnetic fields and temperatures indicated in the legend. The
    magnetic field was applied perpendicular to the film plane.
    For the curves at $B = 0$ and $B = 2$~T data
    for increasing and decreasing voltage are shown in order to demonstrate the reproducibility.
    The nonlinearities observed at low $T$ and low $B$ vanish only at $T > 1.4~$K.}

    \end{center}
\end{figure}

In order to test the interpretation of Coulomb blockade, we record
IVs of tunnel contacts for various magnetic fields, applied
perpendicular to the film plane and for temperatures below and
above $T_c$. Fig. \ref{amg_f5} displays a typical set of IVs,
recorded on sample No. 3 for a contact with a conductance at high
bias voltages of about $0.3~G_0$. At the lowest measuring
temperature and without magnetic field the conductance is reduced
for small voltages $|V| \lesssim 100~\mu$V. For larger voltages a
strong increase is observed. Around $|V| \sim 200~\mu$V the
conductance decreases again. At voltages above $500~\mu$V the IV
is roughly linear. The small gap in the IV at low voltage is
suppressed by applying a magnetic field of only 2~mT, which is
smaller than the critical field of the unpatterned film but could
correspond to the critical field of the lithographiclly patterned
film. We therefore attribute this gap to superconductivity. The
remaining nonlinearities would than have a different origin. The
conductance at intermediate voltages is progressively suppressed
by increasing the field strength. However, the nonlinearities
remain up to fields of $B \simeq 4$~T, gradually evolving from an
enhanced conductance at zero bias to a slightly reduced one.

A possible explanation of the occurrence of such nonlinear IVs
would be a strong enhancement of the critical field of a contact
with high transmission values $\tau \geq 0.9$. A drastically
enhanced critical field was observed previously for atomic-size
contacts of Pb prepared with the help of an STM \cite{suderow} in
a magnetic field applied parallel to the current direction. It was
attributed to the incomplete penetration of the magnetic field in
narrow structures. In our previous measurements on Al contacts and
Au contacts with Al leads fabricated with the same technique and
field direction as the present experiments on Zn contacts, all
signatures of superconductivity vanished at magnetic field values
smaller than or equal to the bulk critical field of Al. However,
in those samples the film structure was optically smooth.
Accordingly, the granularity of the Zn film could be the reason
for the unusually high critical,  the nonlinearities only vanish
at temperatures $ T > 1.4~$K, well above the critical temperature,
which we estimate to be about 750 mK, since the signatures of MAR
for atomic-contacts realized with the same sample disappear at
this temperature.

\begin{figure}[t]
   \begin{center}
\includegraphics[width=0.8\columnwidth,clip]{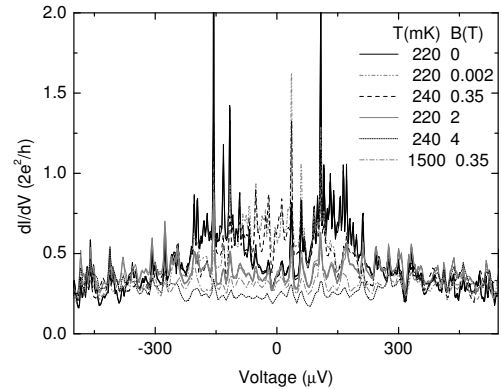}
    \caption{\label{amg_f6} Differential conductance, calculated numerically from the IVs of
 Fig.~5.
    For clarity, the reproducibility is only shown for the curves recorded at
    $B = 0$ and $B = 2~$T.}

    \end{center}
\end{figure}

Therefore and in accordance with the observed shift of the MAR
peaks we attribute the nonlinearities to electron-interaction
effects, such as Coulomb blockade, resonant tunneling or quantum
interference effects. The interplay between electron-electron
interaction effects and superconductivity in atomic-size
structures has so far not been explored theoretically.
Consequently, a detailed interpretation of the observed IVs below
$T_c$ including the magnetic-field behavior is not possible.

Taking the typical voltage scale of the nonlinearities of $|V|
\sim 200~\mu$V as a measure for the charging energy $eV = e^2/2C$,
we deduce a capacity $C = 0.4~$fF corresponding to the capacity of
a sphere with a diameter of 7~$\mu$m. This value is about a factor
10 larger than the typical grain size of the Zn film. Thus, we
interpret our observations to indicate the onset of the
interaction effect. Due to the low-impedance environment of the
circuit and the effective screening of the surrounding metal all
interaction effects would be expected to be damped.

The numerical derivatives $dI/dV$ of the IVs of Fig.\ref{amg_f5},
depicted in Fig. \ref{amg_f6} show small, but reproducible
fluctuations superimposed over the larger scale variations. The
fluctuations have a typical voltage period of $V_c \sim 30~\mu$V,
appear at the same voltage values for both sweep directions (but
different for opposite sign of the voltage) and are correlated
from curve to curve. These findings underline at first that the
fluctuations are not due to experimental noise and second that the
contact has not been changed by varying temperature and magnetic
field. There are several possible origins of the fluctuations,
including (universal) conductance fluctuations (CF) or again the
onset of Coulomb blockade due to the granular structure of the
film. The analysis in terms  of CF yields a typical coherence
length of $L_c = \sqrt{hD/eV_c} \sim 800$~nm roughly corresponding
to the grain size of the evaporated film. A gradual transition
from CF to Coulomb-blockade behavior has recently been observed in
carbon nanotubes \cite{schoenenberger,stojetz}. However, the
physical mechanism giving rise to this transition is not yet
clear.

In conclusion, a single-atom contact of Zn is likely to have a
conductance of $\approx 0.8\thinspace G_0$. The histograms for
both closing and opening the contacts indicate the importance of
different contact geometries. In the IV characteristics in the
superconducting state we find indications of interaction effects,
perhaps due to the granular structure of the evaporated films.
However, due to the lack of systematic measurements and
theoretical investigations the interpretation of the features
being due to a Coulomb-blockade type mechanism remains somewhat
speculative.

\section{Acknowledgments}

We acknowledge enlightening discussions with C. Urbina, D. Esteve
and J. M. van Ruitenbeek. We have enjoyed fruitful interaction
with J. C. Cuevas and we thank him for providing us with his
computer codes. This work was supported by the Deutsche
Forschungsgemeinschaft through SFB 195, SFB 513 and the Alfried
Krupp von Bohlen und Halbach-Stiftung.

\end{document}